\documentclass[aps,pra,twocolumn,groupedaddress, showpacs, showkeys]{revtex4-1}

\usepackage{braket}
\usepackage{amssymb}
\usepackage{amsmath}
\usepackage{graphicx}
\usepackage{float}
\begin{document}

%\preprint{}

\title{Signing Perfect Currency Bonds}


\author{Subhayan Roy Moulick}
\email[]{subhayan@acm.org}
\author{Prasanta K. Panigrahi}
\email[]{pprasanta@iiserkol.ac.in}
%\thanks{}
\affiliation{Indian Institute of Science Education and Research Kolkata, Mohanpur 741246, West Bengal, India}

%\noaffiliation whose secu- rity is based on fundamental principles of quantum physics.

\date{\today}

\begin{abstract}
We propose the idea of a Quantum Cheque Scheme, a cryptographic protocol in which any legitimate client of a trusted bank can issue a cheque, that cannot be counterfeited or altered in anyway, and can be verified by a bank or any of its branches.  We formally define a Quantum Cheque and present the first Unconditionally Secure Quantum Cheque Scheme and show it to be secure against any no-signaling adversary. The proposed Quantum Cheque Scheme can been perceived as the quantum analog of \emph{Electronic Data Interchange}, as an alternate for current \emph{e-Payment Gateways}. 
\end{abstract}

%\keywords{}

\maketitle


\section{Introduction}
Replication of classical information is a significant nuisance in copy-protection. Any physical entity created classically can be, in principle, copied. Currency bonds, printed on textile and paper, are no exception, and any adversary, given sufficient time and resources, can be able to counterfeit currency bonds. However, the quantum regime can circumvent this problem, exploiting the `No Cloning Theorem' \cite{WZ82}, and pave way for unforgeable Quantum Currency that are impossible to counterfeit and can have the property of perfect security. 

The idea of Quantum Money was conceived by Wiesner in 1969 \cite{W83, BBBW83}. While it inspired several fundamental ideas, it did not receive much attention for the next 40 years, possibly due to the limitations of technology. It is only recently that, there has been a surge of interest in the possibility of exploiting the laws of quantum mechanics to create unforgeable tokens for currency. While Wiesner's original scheme was broken recently \cite{L10, MVW13, BNSU14}, the idea of using quantum states to create unforgeable currency persisted. Recent progress in the area have been made by Aaronson \cite{A09}, who formally studied public key quantum money and showed its existence relative to a quantum oracle. In the same paper he also proposed a scheme, without an oracle, based on random stabilizer states. However, it was broken by Lutomirski et al. \cite{LAFGHKS09}, within a year.  Recently Farhi et al. \cite{FGHLS12}, proposed a scheme for quantum money, using ideas from knot theory.  Also Aaronson et al. \cite{AC12}, proposed a scheme for quantum money from hidden subspaces. While the security of Aaronson et al.'s scheme can be proved using a black box security and non-black box security under plausible cryptographic assumptions, the security of Farhi et al.'s scheme is not known and analyzing it would require answering fundamental knot theory problems which has no known practical solutions. 

Another research direction in quantum currency is the invention of Quantum Coins, which aspires anonymity, in addition to security against counterfeiters, owing its origin to the pioneering works of Mosca and Stebila \cite{MS10}. They proposed a scheme based on blind quantum computation that required a verifier to do an obfuscated verification with the bank and learn only the validity of the quantum coin. This however is a private key protocol and requires communication with the bank.

In this paper we propose the idea of Quantum Cheques and present a construction of an Quantum Cheque Scheme with Perfect Security against any No-Signaling adversary.
Generally, in a Quantum Cheque Scheme, a trusted bank acts as a key generation center and provides every account holder with a quantum analogue of a cheque book and can store relevant information about the cheque book secretly. Any account holder, who has a valid 'quantum cheque book' can issue cheques that can be verified by the bank or any of its branches, with which the bank shares a classical communication channel. We present the protocol in an idealized form assuming perfect state preparations, transmissions, and measurements, that can also be realized, efficiently, with few qubit systems, without compromising on the security.

Given the active research with the promise to implement long distance quantum communication networks \cite{BHM96, K08, DM10}, a quantum cheque scheme can be perceived as the quantum analog of \emph{Electronic Data Interchange} (EDI), as an alternate for current \emph{e-Payment Gateways}, used widely in e-commerce, that authorizes credit card payments, which rely on classical communication and computational assumptions for their security. While the present classical protocols, rely on \emph{time-stamping} communications to ensure against double spending, a quantum protocol instinctually averts that problem, due to No-Cloning Theorem. Another major advantage of such quantum cheques will be the fact that they can be realized through physical devices using quantum memories \cite{T01, SAAothers10}, as well as can be used to stream in the quantum internet without the need for quantum memory\cite{MSDHN12}. 
With physical devices, equipped with quantum memories, one can imagine storing quantum states in their computers or smart cards to efficiently perform transactions in person or over the quantum internet. 
While, without Quantum Memory, one would require the protocol to run in \emph{real time}, i.e., the Bank prepares and securely sends the 'quantum cheque book' to the account holder, following which, the account holder (issuer) at once prepares the Quantum Cheque, presents it to the receiver. The receiver then immediately relays the cheque to the Bank, who verifies its (in)validity right away. 

The paper is organized as follows.  Section II discusses the cryptographic tools, required to realize the Quantum Cheque Scheme, namely, Quantum One Way Functions, Swap Testing Circuits, and Digital Signatures. Following that, in Section III, the Quantum Cheques are defined and an Unconditionally Secure Quantum Cheque Scheme is proposed therein. The security of the scheme is analyzed in Section IV. Section V concludes the paper, briefly summarizing the ideas.

%\subsection{Quantum Currency}
%In this section we present the definition of quantum currency, and the goals it is expected to achieve. Intuitively we want the following
%\begin{itemize}
%	\item A trusted Bank (or Mint) must be able to feasibly generate any number of currency.
%	\item A trusted Bank must be able to verify the (in)validity of a currency that has been produced by it.
%	\item Any client must be able compare her/his currency with another and  concur the (in)validity of the currency.  
%	\item No adversary, must be able to forge currency.
%\end{itemize}
%
%\textbf{Definition:} A quantum currency scheme is defined as a \emph{(Gen, Bank, Verify)} tuple, where
%\begin{itemize}
%  	\item{Gen} generates a 
%\end{itemize}	
 
\section{Preliminaries}

\subsection{Quantum One Way Functions: } 
For the present scheme one needs a limited-utility quantum one way function \cite{GD01, CG07P}, based on the fundamental properties of quantum system, where unlike classical bits, qubits can exist in superpositions. An arbitrary quantum state $\Ket{\Phi}$ of a qubit resides in the Hilbert space $\mathbb{C}$ and can be written as, $\Ket{\Phi} = \alpha\Ket{0} + \beta\Ket{1}$, where $\alpha, \beta \in \mathbb{C}$, are the probability amplitudes, satisfying $\lvert \alpha \rvert ^2 +\lvert \beta \rvert ^2 =1$ and $\Ket{0}$ and $\Ket{1}$ form an orthonormal basis.
The distance between two qubit states $\Ket{\phi}$ and $\Ket{\phi'}$ is defined as $\sqrt{1 - \lvert \Braket{\phi \vert \phi'} \rvert^2}$.
Using volumetric analysis, it may be seen that there exists $n$ qubit states $\{ \Ket{\phi_k} ^{\otimes n} \}$, such that  $\Braket{{\phi_k}^{\otimes n} \vert {\phi_{k'}}^{\otimes n} } \leq \delta$ for $k \neq k'$. Buhrman et al.  \cite{BCWD01}, showed for $\delta = 0.9$, the size of the set can be $2^{O(2^n)}$.

A quantum one way function is defined as, 
$$\Psi: k \times \Ket{0}^{\otimes n}  \rightarrow \Ket{\psi_k},$$
where $k \in \{0,1\}^*$ and $\Ket{\psi_k}$ is a $n-$qubit quantum state, such that,
\begin{itemize}
	\item $\Psi$ is easy to compute, i.e., there exists a polynomial-time algorithm that can evaluate $\Psi(k, \Ket{0}^{\otimes n})$ and outputs $\Ket{\psi_k}$,
	\item $\Psi$ is hard to invert, i.e., given $\Ket{\psi_k}$, it is difficult to compute $k$
\end{itemize}

At this point, it may be noted that, this construction can be realized in agreement with Holevo's theorem, which limits the amount of classical information that can be extracted from a quantum state \cite{H77}. For a binary string, $k$, of length $L$, and $C$ copies of $\Ket{\psi_k}$, one can only learn almost $Cn$ bits of information. By having $L>>Cn$, one could achieve a one way function, that is impossible to invert.

\subsection{Fredkin Gate: }
In the classical regime, comparing the equivalence of two bit strings is strightforward, however, due to the no-cloning theorem, $\Psi$ might not be able to produce the exact $\Ket{\psi}$ state. To compare states $\Ket{\psi}$ and $\Ket{\psi'}$, we utilize the Fredkin gate (C-swap gate). We prepare an ancilla qubit $\frac{\Ket{0} + \Ket{1}}{2}$ and perform a controlled swap test on two state $\Ket{\psi}$ and $\Ket{\psi'}$. If $\Ket{\psi} = \Ket{\psi'}$, the ancilla qubit, after performing a Hadamard operator yields $\Ket{0}$, on measuring on a computational basis, and is said to pass the swap test. For $\Braket{\psi \vert \psi'} \leq \delta$, the ancilla qubit, after performing the necessary Hadamard Gates, upon measurement passes the test with probability $\frac{1+\delta^2}{2}$, and fails the test with probability $\frac{1-\delta^2}{2}$. Evidently, the swap test always passes for the same inputs, and sometimes fails if they are different. By repeating the swap test, one can amplify its efficiency.  

\begin{figure}[h!]
\includegraphics{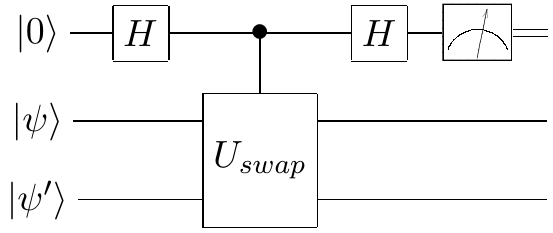}
\caption{depicts a circuit for the Fredkin Gate that non-destructively compares quantum states $\psi$ and $\psi'$, with an additional ancilla qubit}
\end{figure}

\subsection{Digital Signatures: }
A digital signature scheme, $\Pi$, is a 6-tuple $(M, \Sigma, U, Gen, Sign, Vrfy)$, where, 
\begin{itemize}
	\item $M$ is the finite set of valid messages, $\Sigma$ is the finite set of valid signatures, and $U$ is the finite set of users.
	\item The key-generation algorithm, $Gen$, takes in a security parameter $1^k$, and outputs the $Sign, Vrfy$ algorithms and the public parameters. 
	\item The signing algorithm, $Sign$, is a mapping, $Sign: M \times U \rightarrow \Sigma$ 
	\item The verification algorithm, $Vrfy$, is a mapping, $Vrfy: M  \times \Sigma \times U \rightarrow \{True,False\}$.  
\end{itemize}

It is required that, for every $(Sign, Vrfy) \leftarrow Gen(1^k)$, for all $k$, and $m\in M$, and users $i$ and $j$, it holds that
	$$Vrfy_j (m, Sign_i(m), U_i) = True$$

Informally, a digital signature scheme, $\Pi$, must satisfy the following security conditions
\begin{enumerate}
	\item Unforgeability:  Except with a negligible (under a polynomial factor) probability, it should be impossible for an adversary to produce a valid signature
	\item Non-repudiation: Except with a negligible (under a polynomial factor) probability, the signer should not be able disavow a legitimate signature.  
\end{enumerate}

Here, we do not discuss explicit constructions of digital signatures in detail, and instead use an unconditionally secure digital signature scheme, $\Pi = (Gen, Sign, Vrfy)$ as a black box. Suitable constructions of an unconditionally secure classical digital signature using multivariate polynomials have been proposed by Hanoka et al. \cite{HSZI00}, Chuam and Roijakkers \cite{CR90}, where they assume the keys are prepared by a trusted third party. A quantum digital signature scheme have been given by Chuang and Gottesman \cite{GD01, CG07P}.

\section{Quantum Cheques}

\subsection{Definition of a Quantum Cheque Scheme: }
Ideally, a cheque is expected to have the following properties,
\begin{itemize}
	\item A trusted bank or any of its branches must be able to verify the authenticity of a cheque.
	\item An issuer, after issuing a cheque, must not be able to disavow issuing it.
	\item No adversary must be able to \emph{counterfeit} a cheque under some issuer's name or use a cheque more than once to withdraw money.
\end{itemize}

Informally, a Quantum Cheque Scheme consists of three algorithms,
\begin{itemize}
	\item \textbf{Gen}, which takes as input a security parameter and probabilistically generates a 'cheque book' and key for the issuer.  
	\item \textbf{Sign}, which takes as input the issuer's key and amount to be signed, and produces a quantum state $\chi$ called a Cheque. (This state  $\chi$ is an ordered pair $(id, \$, \rho_{\$})$, where $id$ and $\$$ are classical description of the issuer's identity and amount signed respectively, and $\rho_{\$}$ is a quantum cheque state.)   
	\item \textbf{Verify},  which takes in as input the key, and the alleged cheque $\chi$ and decides its (in)validity. 
\end{itemize}

The Scheme is said to have a completeness error $\epsilon$, if $\forall$ valid cheques $\chi$,
	$$Pr \big[ Verify(\chi)\ accepts \big] \geq 1 - \epsilon, $$
	
The Scheme is said to have a soundness error $\delta$, if $\forall$  counterfeiters $\mathcal{C}$, 
	$$Pr \big[ X' \setminus X \neq \varnothing : X' \leftarrow \mathcal{C}( X )  \big] \leq \delta,$$
	
	where $X$ = $\{\chi_1, \chi_2, \cdots ,\chi_q\}$, $\mathcal{C}$ is a counterfeiter that Counterfeits a cheque (formally defined later ) that outputs $X' = \{\chi'_1, \chi'_2, \cdots ,\chi'_{q'}\}$, that \emph{Verify} accepts, and $\varnothing$ denotes an empty set. 

%%% CHANGE formally defined in section IV A to  LATER!!! %%%

\subsection{The Quantum Cheque Scheme: }
For purposes of brevity, we introduce three parties, Alice, Abby and Bank to describe the scheme. The Bank can have several branches and can be thought of as a set of parties connected by a (secure) classical channel with the main branch. The main branch is denoted just as Bank in the rest of the paper. Only the Bank is a trusted party in the protocol, and not necessarily the branches. Alice plays the role of the customer, who issues the cheque to  Abby, the vendor.  Abby then submits it to the Bank (or any of its branches), to encash. The Bank (or any of its branches) verifies the (in)validity of the cheque. In the protocol, we only assume Alice and Bank are honest. Any other player can be dishonest and adversarial.

\textbf{Gen:}
Alice and the Bank create a shared key $k$. This has to be done only once and can be efficiently realized by using, for example, the BB84 Protocol or simply by Alice going to the Bank physically. 

Alice and the Bank also agree on an information theoretically secure digital signature scheme $\Pi = (Gen, Sign, Vrfy)$, and Alice submits her public key, $pk$, to the Bank and secretly stores her Private Key, $sk$.

The Bank prepares a string of $l$ GHZ states,
%$$\big\{ \Ket{\phi^{(i)}}_{GHZ} = \frac{1}{{\sqrt{2}}} \big( \Ket{0^{(i)}}_{A_1}\Ket{0^{(i)}}_{A_2}\Ket{0^{(i)}}_{B}$$
%$$ + \Ket{1^{(i)}}_{A_1^{(i)}}\Ket{1^{(i)}}_{A_2}\Ket{1^{(i)}}_{B} \big) \big\}_{i=0}^l ,$$
\begin{eqnarray*}
\Ket{\phi^{(i)}}_{GHZ} = 	&\frac{1}{{\sqrt{2}}}    &   \big( \Ket{0^{(i)}}_{A_1}\Ket{0^{(i)}}_{A_2}\Ket{0^{(i)}}_{B} 	   \\ 	
					&				& 	+ \Ket{1^{(i)}}_{A_1}\Ket{1^{(i)}}_{A_2}\Ket{1^{(i)}}_{B} \big)   \\
\end{eqnarray*}
with $1 \leq i \leq l$ and corresponding unique serial number $s \in \{0,1\}^n$ and gives two of the three particles (entangled qubits) from every GHZ triplet state and the serial number to Alice (via a secure channel), and stores the third particle (entangled qubit) secretly along with other details in a private database. For conciseness, we adopt the notation $ \{ \Ket{\phi^{(i)}}_{GHZ}  \}_{i=1:l}$ to denote a set(string) of states, $ \{ \Ket{\phi^{(1)}}_{GHZ}, \Ket{\phi^{(2)}}_{GHZ}, \ldots, \Ket{\phi^{(l)}}_{GHZ} \}$, throughout the rest of the paper.

Alice now holds $(id, pk, sk, k, s,  \{\Ket{\phi^{(i)}}_{A_1}, \Ket{\phi^{(i)}}_{A_2} \}_{i=1:l})$.
and the Bank holds $(id, pk, k, s, \{\Ket{\phi^{(i)}}_{B} \}_{i=1:l})$, 

%\emph{Note on GHZ States:} The GHZ states are analogous to classical cheque books issued by a bank. It is possible for Alice and the Bank to have many tuples of GHZ states by simply storing them as $(id, k, s^{(i)}, \Ket{GHZ}_{B}^{(i)})$

\textbf{Sign:}
To Sign a cheque worth amount $M$, Alice generates a random number $r \leftarrow U_{\{0,1\}^L}$ and prepares a $n$-qubit state, 
    $$\Ket{\psi_{alice}}= f(k || id || r || M),$$ 
    
where $f: \{0,1\}^* \times \Ket{0}^{\otimes n} \rightarrow \Ket{\psi}$ is a quantum one way function, $k$ is the secret key otherwise shared only between Alice and Bank, $id$ is the identity of Alice and $x || y$ represents concatenation of two bit-strings $x$ and $y$.

Alice also prepares $l$ states $\{\Ket{\psi_M^{(i)} }\}_{i=1:l}$ corresponding to the amount M using the one way function $g: \{0,1\}^* \times \Ket{0} \rightarrow \Ket{\psi}$, as 
  $$ \Ket{\psi_M^{(i)} } = g(r || M || i) $$
for all $i$, s.t. $1\leq i \leq l$. 
%At this point  Abby, if wishes to, can optionally, also verify the states  $\{\Ket{\psi_M^{(i)} }\}_{i=1:l}$, by computing  $\{\Ket{\psi_M^{,(i)} } \}_{i=1:l} = \{g(r || M || i) \}_{i=1:l}$ and performing a non-destructive swap test using a Fredkin Gate. 

To create the cheque, Alice uses one of her entangled qubits, $\Ket{\phi^{(i)}}_{A_1}$, (with serial number $s$) to encode $\Ket{\psi_M^{(i)} }$ as follows \cite{HBB99}. 

%To encode the $i-$th qubit$, \Ket{\psi^{(i)}}= \alpha \Ket{0} + \beta \Ket{1}$ with the $i-$th GHZ state, Alice combines $\Ket{\psi^{(i)}}$ and one of her entangled qubits from the $i-$th pair, $\Ket{\phi^{(i)}}_{A_1}$, and performs a Bell measurement on the two, causing the system to collapse as $U \otimes I(\alpha \Ket{00} + \beta \Ket{11})_{A_1B}$, where U is one of the Pauli matrices. Alice then performs the suitable gate operations based on the observed Pauli matrix. Now, the information of $\Ket{\psi}$ has been split between $\Ket{\phi^{(i)}}_{A_2}$ and $\Ket{\phi^{(i)}}_{B}$. 
%This encoding procedure is carried out $l$ times for each of the $\{ \Ket{\psi_M ^{(i)} } \}_{i=1:l}$.

To encode the $i-$th qubit$, \Ket{\psi_M^{(i)}}= \alpha_i \Ket{0} + \beta_i \Ket{1}$ with the $i-$th GHZ state, Alice combines $\Ket{\psi_M^{(i)}}$ and one of her entangled qubits from the $i-$th pair, $\Ket{\phi^{(i)}}_{A_1}$, and performs a Bell measurement on the two. More concretely, the four particle entangled state can be described as

\begin{eqnarray}
\label{eq:1}
	\begin{split}
	\Ket{\phi^{(i)}} =		& 	  \Ket{\psi_M^{(i)}} \otimes \Ket{\phi}_{GHZ} \\
			=		&	 \frac{1}{2} \big\{ \Ket{ \Psi^{+}}_{A_1} (\alpha_i \Ket{00}_{A_2B} + \beta_i \Ket{11}_{A_2B} \\
					&	+ \Ket{\Psi^{-}}_{A_1} (\alpha_i \Ket{00}_{A_2B} - \beta_i \Ket{11}_{A_2B} \\	
					&	+ \Ket{\Phi^{+}}_{A_1} (\beta_i \Ket{00}_{A_2B} + \alpha_i \Ket{11}_{A_2B} \\  
					&	+ \Ket{\Phi^{-}}_{A_1} (\beta_i \Ket{00}_{A_2B} - \alpha_i \Ket{11}_{A_2B} \big\}
	\end{split}
\end{eqnarray}

where $\Ket{\Psi^{+}}, \Ket{\Psi^{-}}, \Ket{\Phi^{+}}, \Ket{\Psi^{-}}$ denotes one of the four Bell states, that is then measured by Alice.

If Alice's result, from equation \eqref{eq:1}, is $\Ket{\Psi^+}$ or $\Ket{\Psi^-}$, the Bank's density matrix of its GHZ particle reads
\begin{equation} 
	\label{eq:2}
	\rho_B = \vert \alpha_i \vert^2 \Ket{0}_{BB} \Bra{0} + \vert \beta_i \vert^2 \Ket{1}_{BB} \Bra{1} 
\end{equation}

while if Alice's measurement outcome is $\Ket{\Phi^+}$ or $\Ket{\Phi^-}$, the Bank's density matrix of its GHZ particle reads 
\begin{equation} 
\label{eq:3}
	\rho_B = \vert \beta_i \vert^2 \Ket{0}_{BB} \Bra{0} + \vert \alpha_i \vert^2 \Ket{1}_{BB} \Bra{1}
\end{equation}

Following that, Alice performs a suitable gate operation (Pauli matrix) based on the observed Bell State as follows, 
\begin{center}
	\begin{tabular}{ l c*{4}    l }
		$\Ket{ \Psi^{+}}  \rightarrow I$ 			& & & & &	$\Ket{ \Psi^{-}} \rightarrow \sigma_Z$ 	\\
		$\Ket{ \Phi^{+}} \rightarrow \sigma_X$	& & & & &	$\Ket{ \Phi^{-}} \rightarrow \sigma_Y$	\\
	\end{tabular}
\end{center}

Now, the information of $\Ket{\psi^{(i)}}$ has been split between $\Ket{\phi^{(i)}}_{A_2}$ and $\Ket{\phi^{(i)}}_{B}$. 
Based on the observed Bell State, Alice performs a suitable error correction (Pauli Matrix) on $\Ket{\phi^{(i)}}_{A_2}$, that she posses. This encoding procedure is carried out $l$ times for each of the $\{ \Ket{\psi_M ^{(i)} } \}_{i=1:l}$.

Alice also signs the serial number $s$ as $\sigma \leftarrow Sign_{sk}(s)$.

Alice finally produces a Quantum Cheque $$\chi = (id, s,  r, \sigma, M, \{\Ket{\phi^{(i)}}_{A_2}\}_{i=1:l}, \Ket{\psi_{alice}} )$$ and gives it to  Abby.

\begin{figure*}
\includegraphics[scale=0.45]{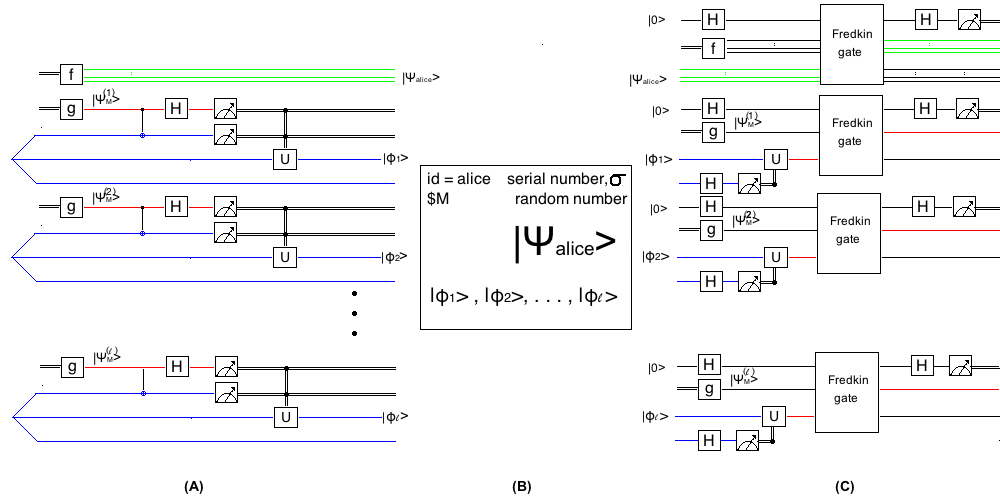}
\caption{(A) Circuit used for signing a Quantum Cheque by Alice, (B) The Quantum Cheque, (C) Circuit used for verifying the Quantum Cheque by the Bank}
\end{figure*}

\textbf{Verify:}
 Abby when produces the Quantum Cheque $\chi = (id, s,  r, \sigma, M, \{\Ket{\phi^{(i)}}_{A_2}\}_{i=1:l}, \Ket{\psi_{alice}} )$ at any of the valid branches of the bank, the branch communicates (securely) with the Bank's main branch, and checks the validity of the $(id,s)$ pair and runs a verification using $Vrfy_{pk}(\sigma, s)$. If $(id,s)$ and $\sigma$ is invalid, the branch destroys the cheque and aborts. Else, the respective branch continues with the verification.
 
%The main branch now performs a measurement, in the Hadamard basis, on its copy of   $\Ket{\phi}_{B}$, and communicates (securely) the results via a classical channel to the appropriate Branch, which then uses the $\Ket{\phi}_{A_1}$, to recover $\{ \Ket{\psi_M ^{(i)}} \}_{i=1:l}$. 
%It computes $\Ket{\psi_M^{,(i)}}  = f(M || i)$ and uses a Fredkin Gate to perform a non-destructive swap test on the states $\Ket{\psi_M^{(i)}}$ and $\Ket{\psi_M^{,(i)}}$ for all $i=1:l$. 

The main branch now performs a measurement, in the Hadamard basis, on its copy of   $\Ket{\phi}_{B}$, to obtain outcomes $\Ket{+}$ or $\Ket{-}$ and communicates (securely) the results via a classical channel to the appropriate Branch. Based on the outcome, the Branch performs the following Pauli Matrix on $\Ket{\phi^{(i)}}_{A_2}$, to recover $\Ket{\psi_M ^{(i)}} $. 
\begin{center}
	\begin{tabular}{ l c*{4}    l }
		$\Ket{ +}  \rightarrow I$ 			& & & & &	$\Ket{ -} \rightarrow \sigma_Z$ 	\\
	\end{tabular}
\end{center}

This is done $l$ times for each of  $\{  \Ket{\phi^{(i)}}_{A_2} \}_{i=1:l}$, to recover $\{ \Ket{\psi_M ^{(i)}} \}_{i=1:l}$. 
The Bank computes $\{\Ket{\psi_M^{,(i)} } \}_{i=1:l} = \{g(r || M || i) \}_{i=1:l}$ and performs a swap test on each state $\{\Ket{\psi_M^{,(i)} } \}$ and $\{\Ket{\psi_M^{(i)} } \}$.

The Bank also computes $\Ket{\psi'_{alice}} = f(k || id || r || M)$ and again performs a non-destructive swap test on states $\Ket{\psi_{alice}}$ and $\Ket{\psi'_{alice}}$. 

The Bank (or branch) accepts the cheque if both the swap tests pass, i.e., if $\Braket{\psi_{alice} \vert \psi'_{alice}} \geq \kappa_1$ and $\{ \Braket{\psi_M^{(i)} \vert \psi_M^{,(i)}}  \geq \kappa_2 \}_{i=1:l}$, where $\kappa_1$ and $\kappa_2$ are the thresholding constants, that serve as security parameters determined by the bank. The Branch rejects and aborts the transaction otherwise, and also destroys the cheque.

\section{Security of the Quantum Cheque Scheme}

\subsection{Impossibility of Counterfeiting: }
For purposes of contradiction, suppose there exists an adversary $\mathcal{A}$ that breaks the proposed Quantum Cheque Scheme, X.
Let $\chi \leftarrow Expt_{\mathcal{A},X}(1^k)$ denote the experiment,
$$ (params) \leftarrow Gen(1^k), \chi \leftarrow \mathcal{A}^{Sign( \cdot )}(id, pk_{id})$$
where $params$ are the parameters generated by the algorithm $Gen( \cdot )$, $k$ the security parameter and $\mathcal{A}$ is allowed polynomially bounded number of queries to its signing oracle, $Sign(\cdot)$, for a signer $id$. Let $\{M_1, M_2 \ldots M_q\}$ be the amounts $\mathcal{A}$ queries the signing oracle in a particular experiment, to get Quantum Cheques $\{\chi_1,\chi_2, \ldots, \chi_q\}$ respectively.

Let Counterfeit be the event, 
$$ (Verify (\chi') = 1) \wedge \chi' \notin {\chi_1,\chi_2, \ldots, \chi_q} $$

We define,
$$ Succ_{\mathcal{A}, X} = Pr \big[  \chi' \leftarrow Expt_{\mathcal{A}, X}(1^k) : Counterfeit \big]$$

For the scheme $X$ to be unconditionally secure, $Succ_{\mathcal{A}, X}$ must be negligible for any adversary, $\mathcal{A}$, with unbounded computational resources and time where $\mathcal{A}$ is only limited by physical laws.

An unknown quantum state $\Ket{\psi}$ cannot be perfectly cloned, as known from the no cloning theorem.  
Imperfect cloning as shown by Bu{\v{z}}ek and Hillary  \cite{BH96}, where an algorithm that takes $\psi$ as input and outputs two qubits such that the reduced density matrix of either output qubit, $\rho$, satisfies $\Braket{\psi \vert \rho \vert \psi} =5/6$. This was later proved to be optimal \cite{GM97, BDECS98}.  By having the security parameter $\kappa$, used by the bank for the swap tests, such that $\kappa > 0.91$, it can be made impossible for any adversary to copy states and reuse the same cheque more than once.

It must be noted, an adversary, that successfully forges a new cheque for an $id$, would require the knowledge of the (classical) secret key, $k$, corresponding to $id$. However, even if the adversary has access to $q$ different cheques with $n$ qubits per cheque, signed by a key $k \in \{0,1\}^L$, Holevo's theorem limits the amount of classical information that can be extracted from a quantum state \cite{H77}. By having $L>>qn$, it can be made impossible for an adversary to extract relevant information about the secret key, and can only guess the key with probability $P \leq 2^{-(O(L)-qn)}$. 

Also, the fact that the adversary does not have access to the required pairs of GHZ states, that are otherwise shared only among a trusted Bank and the issuer, it is impossible for an adversary to produce a cheque $\chi$ that would be verified by a Bank, due to the fact that entanglement is monogamous \cite{FLV88, W89, CKW00, HSS03} and it would be \emph{impossible} to produce states $\{\Ket{\phi_M ^{(i)} }\}_{i=1:l}$. This can be traced back to the unconditional security of Hillary et al.'s \cite{HBB99} seminal work on quantum secret sharing.

%Also given that no adversary has access to any of the entangled qubits of the GHZ states that are otherwise shared only among a trusted Bank and the issuer, it is impossible for an adversary to produce a cheque $\chi$ that can be verified by a bank, due to the fact that entanglement is monogamous \cite{FLV88, W89, CKW00, HSS03} and it would be impossible to produce states $\{\Ket{\phi_M ^{(i)} }\}_{i=1:l}$ and guess the corresponding serial number $s$. This can be traced back to the unconditional security of Hillary et al.'s \cite{HBB99} seminal work on quantum secret sharing.

The only remaining strategy for an adversary, that has access to $q$ cheques $\{ \chi_1, \chi_2 \ldots \chi_q \}$, is to perform some unitary operation to at least one of the cheques, $\chi_j$ and modify it and produce a tuple $\chi' =  (id, s', r', \sigma', M', \{ \Ket{ \phi^{,(i)} }_{i=1:l} \}, \Ket{\psi'_{Alice}} )$, such that $\chi' \notin \{ \chi_1, \chi_2 \ldots \chi_q \}$ and $Verify$ accepts $\chi'$, where $M' \neq M$ and $\exists j$, s.t., $\Ket{ \phi^{,(j)} } \neq \Ket{ \phi^{(j)} }$ and $\Ket{\psi'_{Alice}} \neq \Ket{\psi_{Alice}}$

For purposes of contradiction, suppose the adversary, $\mathcal{A}$ successfully modifies a cheque $\chi = (id, s, r, \sigma, M, \{ \Ket{ \phi^{(i)} }_{i=1:l} \}, \Ket{\psi_{Alice}} )$ to produce $\chi' = (id, s', r', \sigma', M', \{ \Ket{ \phi^{,(i)} }_{i=1:l} \}, \Ket{\psi'_{Alice}})$. 

Basically since he cannot forge new cheques (due to reasons mentioned earlier), he can only manipulate the qubits stored in the different registers of the quantum cheques he has access to, up to an unbounded number of Unitary operations. Clearly if $M' \neq M$ or $r' \neq r$, then the adversary at least needs to produce a corresponding \emph{signature state} $\Ket{\psi_{alice}}$, which in turn would imply the Quantum One Way Function is not secure, and that happens w.p. $\leq 2^{-(O(L)-qn)}$. 
Another strategy for an adversary would be to not modify the \emph{signature state}, but the \emph{entangled qubits} using only local operations on the adversaries local system.

However, it can be seen for an entangled state, say $\Ket{\Psi} = \alpha \Ket{00} + \beta \Ket{11}$ represented by $\rho = \Ket{\Psi} \Bra{\Psi}$, and given access to only one qubit, $\rho_a = Tr_b(\rho_{ab})$ , if it were possible to modify that to produce $\Ket{\Psi'} = \alpha' \Ket{00} + \beta' \Ket{11}$ for a specific value of $(\alpha' , \beta')$, it would imply signaling. This is because the adversary can do a local operations and set the values of $\alpha'$ and $\beta'$ to $\sqrt{1-\epsilon}$ and $\sqrt{\epsilon}$ (or vice versa) to send a message bit $0$ (or $1$), with a party with whom he shares entanglement with, faster than the speed of light. 

So, any $\mathcal{A}$ that can Counterfeit, can violate the No-Signaling Principle or breaks the QOWF. Hence as long as the No-Signaling Principle holds, the $Succ_{\mathcal{A}, X}$ must be $\leq 2^{-(O(L)-qn)}$.

\subsection{Impossibility of Non-Repudiation by Signatory: }
For purposes of contradiction, suppose there exists an issuer, $\mathcal{A}$, which can repudiate a quantum cheque issued by it earlier. Then we can construct an algorithm $\mathcal{B}$ that can break the unconditionally secure digital signature scheme $\Pi$ and violate the property of nonrepudiation. 

This is straightforward to see, where $\mathcal{B}$ allows $\mathcal{A}$ to access the Sign Oracle and after polynomially many queries, $\mathcal{B}$ requests $\mathcal{A}$ a cheque $\chi= (id, s,  r, \sigma, M, \{\Ket{\phi}_{\mathcal{A}_2}\}_{i=1:l}, \Ket{\psi_{\mathcal{A}}} )$ and a (possibly zero knowledge) proof, $P$, of $\mathcal{A}$'s capacity to repudiate the cheque. $\mathcal{B}$ then simply produces $(s, \sigma)$ as a signature and the proof, $P$, as a challenge to the property of the non-repudiation of the claimed Unconditionally Secure Signature (USS) Scheme, $\Pi$. 
However, this leads to a contradiction to the assumption that $\Pi$ is an unconditionally secure scheme.

Here only a proof sketch is given. To analyze the security of the scheme to its full glory, is left for future study \cite{SP00, N09}.

\section{Conclusion}
To conclude, we have put forward here, the idea of utilizing quantum states to fabricate currency bonds, in form of quantum cheques and presented the first construction for an unconditionally secure quantum cheque scheme, where a bank and its client share GHZ states and a classical bit string as a secret key. The client can issue quantum cheques, $\chi$, and these quantum states can be physically stored in a quantum memory or transmitted in the anticipated quantum internet without the need for long term storage of qubits. The bank, or any of its branches that are connected by a classical channel, can verify the (in)validity of an alleged quantum cheque. 
The proposed scheme is claimed to be unconditionally secure based on fundamental laws of quantum information - Holevo's Theorem and the No-Signaling Theorem. %and the fact that no adversary, without the entangled GHZ states, would be able to counterfeit a quantum cheque. 
Also the fact that it is impossible to clone quantum states prevents replication of quantum cheques for a utility gain of an adversary.
Such cheques are expected to play a pivotal role in the much anticipated quantum internet as payment gateways and can also be used in a consumerist market where the quantum states can be stored in quantum memories.

\begin{acknowledgments}
S.R.M. thanks Robin Kothari (MIT) for his helpful references to the literature of imperfect cloning and Madhur Srivastava (Cornell) and Anirban Dey for helpful discussions.
\end{acknowledgments}

\bibliography{references}

\end{document}